\documentclass[aps,prl,twocolumn,superscriptaddress]{revtex4-2}
\usepackage{bbm}
\usepackage{graphicx}
\usepackage{dcolumn}
\usepackage{bm}
\usepackage{subfigure}
\usepackage{amsmath}
\usepackage{feynmf}
\usepackage{hyperref}
\usepackage{CJK}
\usepackage{amssymb}
\usepackage{attachfile}
\newcommand{\bk}{\boldsymbol k}

\newcommand{\bA}{\boldsymbol{A}}

\newcommand{\bn}{\boldsymbol{n}}
\newcommand{\zb}{\color {black}}

\usepackage{braket}
\usepackage{esint}
\usepackage{times}

\begin{document}

\title{Floquet Weyl Semimetals with Linked Fermi Arcs}

\author{Dongling Liu}
\affiliation{Guangdong Provincial Key Laboratory of Magnetoelectric Physics and Devices,
State Key Laboratory of Optoelectronic Materials and Technologies,
School of Physics, Sun Yat-sen University, Guangzhou 510275, China}

\author{Zheng-Yang Zhuang}
\affiliation{Guangdong Provincial Key Laboratory of Magnetoelectric Physics and Devices,
State Key Laboratory of Optoelectronic Materials and Technologies,
School of Physics, Sun Yat-sen University, Guangzhou 510275, China}

\author{Zhongbo Yan}
\email{yanzhb5@mail.sysu.edu.cn}
\affiliation{Guangdong Provincial Key Laboratory of Magnetoelectric Physics and Devices,
State Key Laboratory of Optoelectronic Materials and Technologies,
School of Physics, Sun Yat-sen University, Guangzhou 510275, China}

\date{\today}

\begin{abstract}
Floquet engineering provides a powerful and flexible method for modifying the 
band structures of quantum materials. While circularly polarized light  has been 
shown to convert curved nodal lines in three-dimensional semimetals into Weyl points, 
such a transformation is forbidden for an isolated straight nodal line. In this work, 
we uncover a dramatic shift in this paradigm when multiple straight nodal lines intersect. 
We observe that circularly polarized light not only gaps them into Weyl points but also 
induces unprecedented surface-state Fermi arcs that extend across the entire surface Brillouin 
zone and form a linked topological structure. These findings advance our fundamental understanding 
of light-driven transitions in topological semimetals and unveil a unique Weyl semimetal phase 
defined by linked Fermi arcs. We discuss potential exotic phenomena arising from this phase, applications of 
our predictions to spin-split antiferromagnets, and the extension of this Weyl semimetal phase to classical systems. 

\end{abstract}

\maketitle

{\it Introduction.---}A groundbreaking advance in Floquet engineering emerged
 with the discovery that circularly polarized light (CPL) can open a band gap at graphene's Dirac points~\cite{Oka2009}, 
realizing topological gapped phases now known as Floquet Chern insulators.
This seminal work ignited widespread investigation into Floquet topological phases~\cite{Kitagawa2010b,lindner2011floquet,Jiang2011,rudner2013anomalous,Grushin2014,Potter2016,Harper2017,Roy2016periodic,Yao2017floquet,Morimoto2017floquet} 
and light-driven topological phase transitions in quantum materials~\cite{Oka2019review,Rudner2020review,Bao2022review,Zhan2024review}.
The direct observation of light-driven band structure modifications, however, 
was challenging for years until recent notable advances in 
time- and angle-resolved photoemission spectroscopy (TrARPES).
This technique's unprecedented capability to resolve photon-dressed band structure in momentum space~\cite{wang2013observation,Zhou2023Floquet,Choi2025,Merboldt2025,Bielinski2025} 
is now propelling the field into an exciting new phase of discovery.

\begin{figure}[t]
\centering
\includegraphics[width=0.45\textwidth]{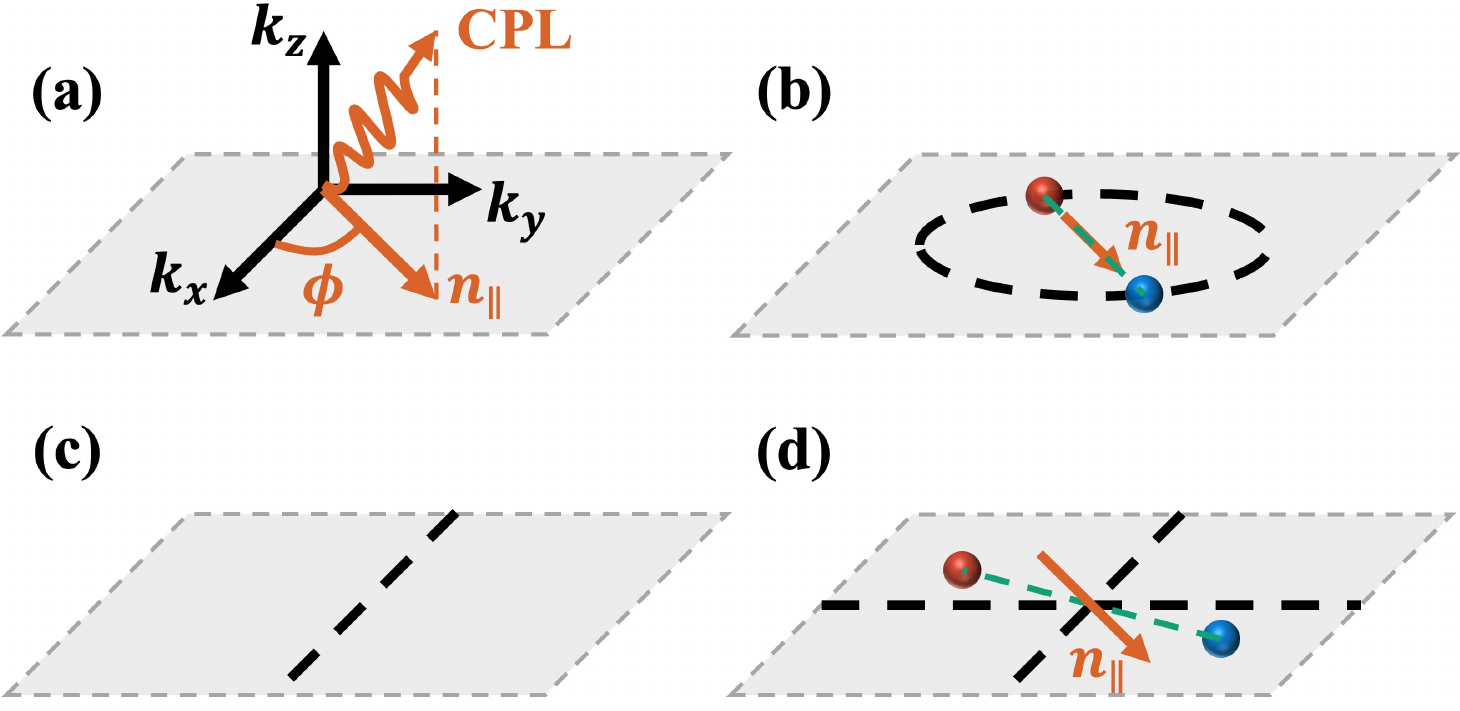}
\caption{Schematic diagram illustrating the influence of CPL on different types of nodal lines.
Dashed black lines denote gapped nodal lines, while solid dots represent CPL-induced FWPs of opposite chirality (red/blue).
 (a) CPL is incident along a general direction.  
 (b) Two FWPs emerge from a circular nodal ring, displaced along $\bn_{\parallel}$ (light’s in-plane projection). 
 (c) No FWPs emerge for an isolated SNL. (d) Two FWPs appear near two SNLs’ crossing point, with 
 their displacement vector generally non-parallel to $\bn_{\parallel}$. 
}\label{fig1}
\end{figure}

Band degeneracy serves as the source for the nontrivial 
topological properties of the band structure~\cite{Chiu2015RMP}. The influence of CPL 
on band degeneracy depends critically 
on both the system's dimension and the structure of the degeneracy. In two-dimensional (2D) systems, 
the most common type of band degeneracy is Dirac points protected by $\mathcal{PT}$ symmetry. 
Since CPL generally breaks $\mathcal{PT}$ symmetry, 2D Dirac points are unstable under CPL irradiation. 
Consequently, the 2D Floquet topological phases induced by CPL typically 
exhibit gapped band structures~\cite{cayssol2013floquet,Usaj2014,Wang2018floquet}. In three-dimensional (3D) systems, band degeneracies exhibit 
greater diversity, encompassing nodal points~\cite{wan2011,Xu2011,wang2012dirac,wang2013three,Bradlyn2016}, 
nodal lines~\cite{Burkov2011nlsm,Fang2015nodal,Bzdusek2016,Yan2018chain,Chen2017link,Yan2017link,Ren2017knot,Ezawa2017link,Chang2017link}, and nodal surfaces~\cite{Liang2016NS,Wu2018NS}. 
This diversity leads to a richer interplay between CPL and band degeneracy compared to 2D systems.
Among these band degeneracies, Weyl points are special because
their stability is topologically protected~\cite{Armitage2018RMP}, 
unlike other band degeneracies that rely on symmetry protection. 
When subjected to CPL, Weyl points primarily experience chirality-dependent positional shifts~\cite{Chan2016hall,Day2024},
while other degeneracies typically transform into Weyl points~\cite{Hubener2017,Bucciantini2017,yan2016tunable,Chan2016type,Narayan2016,Taguchi2016,Yan2017Weyl,Ezawa2017FWSM,Chen2019FWSM,Du2022FWSM,Liu2025FWSM,
Zhang2018FWSM,Ghorashi2018FWSM,Liu2018FWSM,Trevisan2022,Wang2023FWSM,Yang2024FWSM,Huang2024FWSM,Pandit2025,Hirai2024,Fan2024}. 
This conversion results in a robust topological gapless phase---the Floquet Weyl semimetal.
Particularly interesting is the transition from curved nodal lines to Weyl points [illustrated in Fig.~\ref{fig1}(b)], 
which typically generates widely separated Weyl points~\cite{yan2016tunable,Chan2016type,Narayan2016} and enables the realization 
of multi-Weyl points~\cite{Yan2017Weyl,Ezawa2017FWSM}. This large separation 
leads to extended Fermi-arc surface states at the boundary and gives 
rise to pronounced anomalous Hall effects.

Crystalline symmetries can enforce nodal lines to be straight~\cite{Li2017SNL,Liu2021SNL,Wu2021SNL,Wang2022SNL,Fernandes2024,He2024SNL,Tang2024,Wang2025SNL,Zhuang2025SNL}. 
However, research on Floquet Weyl points (FWPs) arising 
from nodal lines has thus far been confined to curved cases~\cite{yan2016tunable,Chan2016type,Narayan2016,Taguchi2016,Yan2017Weyl,Ezawa2017FWSM,Chen2019FWSM,Du2022FWSM,Liu2025FWSM}. 
This restriction stems from the 
fact that a straight nodal line (SNL) effectively represents a stack of 2D Dirac 
points along a specific momentum direction. Since these Dirac points are identical, CPL
affects them uniformly---either gapping all of them simultaneously or leaving the entire stack unaffected (as 
demonstrated below). 
Consequently, this uniform response prevents the generation of FWPs from SNLs [illustrated in Fig.~\ref{fig1}(c)].
In this work, we demonstrate that this paradigm undergoes a dramatic transformation when multiple SNLs intersect. 
Beyond the unexpected emergence of FWPs, we discover an extraordinary characteristic of 
the resulting Fermi-arc surface states: they span the entire surface Brillouin zone (BZ), 
forming an intricately linked structure---a phenomenon never before observed in either static or Floquet Weyl semimetals. 
We also investigate the anomalous Hall effect in this system, 
uncovering a strong dependence on the propagation direction of the incident light.

{\it A single SNL under CPL irradiation.---}To appreciate the striking difference 
between isolated and intersected SNLs, we first show
the effect of CPL on an isolated SNL. Specially, we consider the following continuum model:
\begin{eqnarray}
\mathcal{H}(\bk)=k_{y}\sigma_{x}+k_{z}\sigma_{y},
\end{eqnarray}
where $\bk=(k_{x},k_{y},k_{z})$. The energy spectra are given by $E_{\pm}(\bk)=\pm\sqrt{k_{y}^{2}+k_{z}^{2}}$, featuring
a SNL along the $k_{x}$ axis.
We consider that CPL is incident in the direction $\bn = (\cos\phi \sin \theta,\sin\phi \sin \theta,\cos \theta)$,
where $\phi$ and $\theta$ are the azimuthal and polar angles in the
spherical coordinate system [illustrated in Fig.~\ref{fig1}(a)], respectively.
The vector potential of the light is described by $\bA(t) = A_{0}[\cos(\omega t)\mathbf{e}_{1} + \eta \sin(\omega t)\mathbf{e}_{2}]$.
Here, $\omega$ is the angular frequency and $\eta=\pm1$ labels the two types of handedness. The two unit vectors, 
$\mathbf{e}_{1} = (\sin \phi, -\cos\phi,0)$ and
$\mathbf{e}_{2} = (\cos \phi \cos\theta,\sin\phi\cos\theta,-\sin\theta)$, are orthogonal and lie
perpendicular to $\bn$ due to the light's transverse-wave nature. 

The effect of CPL is incorporated through the minimal coupling, $\mathcal{H}(\bk)\rightarrow \mathcal{H}(\bk+e\bA)$, 
where $e$ denotes the elementary charge and we take $e>0$. We focus on the high-frequency off-resonant 
regime. According to the Floquet-Magnus theory, the system is effectively described by a time-independent Hamiltonian 
that is given by~\cite{Kitagawa2011Floquet,Goldman2014} 
\begin{eqnarray}
\mathcal{H}_{\rm eff}(\bk)&=&\mathcal{H}_{0}(\bk)+\sum_{n\geq1}\frac{[\mathcal{H}_{+n},\mathcal{H}_{-n}]}{n\omega}+O(\omega^{-2}),
\end{eqnarray}
where $\mathcal{H}_{n}$ denotes the $n$th-decomposition coefficient matrix, i.e., 
$\mathcal{H}(\bk,t)=\sum_{n}\mathcal{H}_{n}(\bk)e^{in\omega t}$ with $n\in \mathbb{Z}$. For this Hamiltonian, 
we have $\mathcal{H}_{0}(\bk)=k_{y}\sigma_{x}+k_{z}\sigma_{y}$
and $\mathcal{H}_{\pm 1}(\bk)=-\frac{eA_{0}}{2}[
(\cos\phi\pm i\eta\sin\phi\cos\theta)\sigma_{x}\mp i\eta\sin\theta\sigma_{y}]$,
which yields:
\begin{eqnarray}
\mathcal{H}_{\rm eff}(\bk)&=&k_{y}\sigma_{x}+k_{z}\sigma_{y}-\eta\frac{(eA_{0})^{2}}{2\omega}\cos\phi\sin\theta\sigma_{z}.
\end{eqnarray}
It is readily seen that the SNL remains gapless only when the light is incident perpendicular to the SNL, 
corresponding to either $\theta=\{0,\pi\}$ or $\phi=\{\frac{\pi}{2},\frac{3\pi}{2}\}$. In all other orientations, 
the SNL develops a gap. Notably, in contrast to curved nodal lines~\cite{yan2016tunable,Chan2016type,Narayan2016}, 
FWPs never emerge in this system under any illumination conditions. 

Although FWPs cannot be generated, fully gapping out the SNL instead results in a 3D Floquet Chern insulator characterized by 
layer Chern numbers $(C(k_x), C(k_y), C(k_z))$.
 Here, $C(k_i)$ denotes the layer Chern number defined on the 2D momentum plane with fixed $k_i$, given explicitly by $C(k_{i})=\frac{1}{2\pi}\sum_{j,l}\epsilon_{ijl}\int dk_{j}dk_{l}\partial_{j}A_{l}$
where $\epsilon_{ijl}$ is the third-order antisymmetry tensor, 
and $A_{l}=i\langle u(\bk)|\partial_{l} u(\bk)\rangle$ is the Berry connection of the occupied band. 
For the continuum Hamiltonian considered here, we find
\begin{eqnarray}
(C(k_{x}),C(k_{y}), C(k_{z}))=(\frac{\eta\text{sgn}(\cos\phi)}{2},0,0).
\end{eqnarray}
The half-integer value arises from the non-compact nature of the momentum space (every momentum plane belongs to $\mathbb{R}^{2}$). 
However, for lattice Hamiltonians where the BZ is a closed torus, 
the Chern number must be quantized to integers. This implies that SNLs with linear dispersion 
must appear in pairs in lattice systems---a direct consequence of the fermion doubling theorem~\cite{NIELSEN1981a}.

{\it Two intersected SNLs under CPL irradition.---}We now show how intersected SNLs lead to 
qualitatively new behavior. As the simplest example, we analyze two 
intersected SNLs, described by the Hamiltonian
\begin{eqnarray}
\mathcal{H}(\bk)=k_{z}\sigma_{x}+k_{x}k_{y}\sigma_{y}.\label{CSNL}
\end{eqnarray}
The two SNLs are located along the $k_{x}$ and $k_{y}$ axes 
and intersect at $\bk=(0,0,0)$. 
Applying the Floquet-Magnus theory yields
\begin{eqnarray}
\mathcal{H}_{\rm eff}(\bk)&=&k_{z}\sigma_{x}+(k_{x}k_{y}-\frac{(eA_{0})^{2}}{4}\sin2\phi\sin^{2}\theta)\sigma_{y}\nonumber\\
&&+\eta\frac{(eA_{0})^{2}}{2\omega}\sin\theta(-\cos\phi k_{x}+\sin\phi k_{y})\sigma_{z}.
\end{eqnarray}
Depending on the direction of light, the fate of the two SNLs can be categorized into three cases: 
(i) $\theta=0$ or $\pi$. The SNLs remain intact, as all driving-induced terms vanish. 
This aligns with the conclusion that an SNL is stable against
CPL incident perpendicular to it. 
(ii) $\theta\neq0$ or $\pi$, with $\phi= p\pi/2$ ($p\in\{0,1,2,3\}$). One 
SNL becomes gapped, while the other remains gapless. (iii) When $\theta\neq0$ or $\pi$, 
and $\phi\neq p\pi/2$. In these general cases, 
both SNLs become gapped while 
two FWPs emerge [illustrated in Fig.~\ref{fig1}(d)], positioned at 
\begin{eqnarray}
\bk_{w}=\pm\frac{(eA_{0})\sin\theta}{\sqrt{2}}(\sin\phi,\cos\phi,0).\label{Weyl}
\end{eqnarray}
The emergence of FWPs reveals that intersected SNLs host 
fundamentally distinct physical behavior compared to their isolated counterparts. 
{\zb In Eq. (\ref{Weyl}), a notable property is that the positions of FWPs depend solely on the amplitude $A_{0}$ 
and are independent of the frequency $\omega$. This independence implies a remarkable robustness: the high-frequency 
expansion accurately predicts the Weyl-point positions even when extended into lower-frequency regimes where 
it fails to capture the full band structure. }

The FWPs emerging from intersected SNLs exhibit several striking differences compared to those generated in other systems. 
Specifically, we observe that: 
(i) A distinctive scaling of Weyl-point separation ($d_{w}$). 
For FWPs originating from a Dirac point, the separation scales as
 $d_{w}\propto A_{0}^{2}/\omega$~\cite{Hubener2017}. {\zb Such a scaling 
 is also applied to other nodal point semimetals~\cite{Fan2024}. }
 For a circular nodal ring, 
 $d_{w}$ is approximately equal to the ring’s diameter [illustrated in Fig.~\ref{fig1}(b)] and 
 thus independent of $A_{0}$ and $\omega$ to the leading order~\cite{yan2016tunable,Chan2016type,Narayan2016}; 
 In the present case, the scaling follows 
$d_{w}\propto A_{0}$, distinguishing it from both Dirac points and nodal rings. (ii) A distinctive dependence of 
the displacement vector of FWPs on the light's direction. 
When the light propagates along $(\phi,\theta)$ on the Bloch sphere, 
the displacement vector aligns with the light direction for the case of Dirac points~\cite{Hubener2017}. 
For a circular nodal ring, it aligns with  $(\phi,\frac{\pi}{2})$, corresponding to  the in-plane projection of the light's orientation~\cite{yan2016tunable,Chan2016type,Narayan2016}, as illustrated in Fig.~\ref{fig1}(b).
In stark contrast, our analysis [Eq.~(\ref{Weyl})] reveals that for intersected SNLs, the vector is oriented along 
$(\frac{\pi}{2}-\phi,\frac{\pi}{2})$, which generally deviates from the light's orientation or its projection, 
as illustrated in Fig.~\ref{fig1}(d). 

\begin{figure}[t]
\centering
\includegraphics[width=0.45\textwidth]{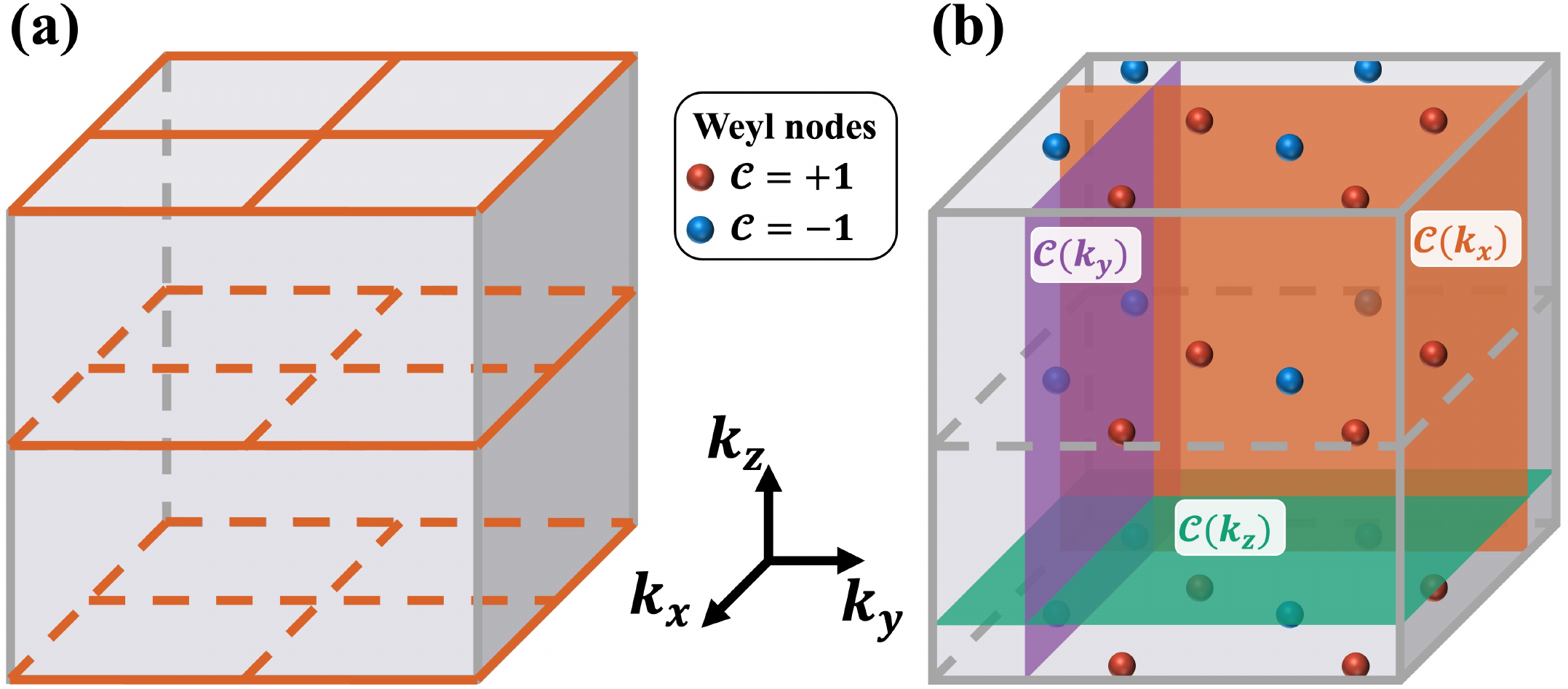}
\caption{ (a) Red lines mark the locations of the SNLs before driving. (b) Configuration of 
the emergent FWPs in the driven system, with red and blue dots denoting 
FWPs of opposite chirality. Note that these FWPs at the top and bottom 
surfaces belong to the same group due to BZ periodicity. The layer Chern numbers are 
defined in these gapped planes (color-coded) away from the FWPs.  
}\label{fig2}
\end{figure}

{\it Linked Fermi arcs.---}A defining universal feature of Weyl semimetals 
is the existence of Fermi arcs in the surface BZ---topological surface states 
that connect the projections of bulk Weyl points, as required by the bulk-boundary correspondence~\cite{Armitage2018RMP}. 
Although our continuum Hamiltonian analysis shows that FWPs can emerge from intersected SNLs, 
determining the connectivity pattern of Fermi arcs and Weyl-point projections requires a lattice-regularized description. 
We achieve this through the standard lattice 
regularization substitution $k_{i}\rightarrow\sin k_{i}$, which yields the following lattice Hamiltonian:
\begin{eqnarray}
\mathcal{H}(\bk)=\sin k_{z}\sigma_{x}+\sin k_{x}\sin k_{y}\sigma_{y}.
\end{eqnarray}
For notational simplicity, we have set the lattice constants to unity. 
The resulting lattice Hamiltonian supports eight SNLs along high-symmetry lines in the bulk BZ: four in the
 $k_{z}=0$ plane and four in the $k_{z}=\pi$ plane, as 
illustrated in Fig.~\ref{fig2}(a). Near each of the eight time-reversal invariant momenta where two SNLs intersect, 
a low-energy expansion yields a continuum Hamiltonian identical in form to Eq. (\ref{CSNL}), 
modulo sign differences in the coefficients. This implies that eight pairs of FWPs
 will emerge under general light irradiation. To verify this prediction, we follow 
the same analytical procedure and derive the Floquet Hamiltonian, which reads
\begin{eqnarray}
\mathcal{H}_{\rm eff}(\bk)=C\sin k_z \sigma_x+D(\mathbf{k}_s)\cos k_z \sigma_z+F(\mathbf{k}_s)\sigma_{y},\label{latticeH}
\end{eqnarray}
where $\mathbf{k}_s=(k_{x},k_{y})$, $C=J_0(\lambda_3eA_0)$, 
$F(\mathbf{k}_s)=-\frac{1}{2}\left[J_0(\lambda_1eA_0)\cos k_+-J_0(\lambda_2eA_0)\cos k_-\right]$, 
and 
\begin{eqnarray}
D(\mathbf{k}_s)&=&-2\eta \frac{J_1(\lambda_3eA_0)}{\omega}\left[\sin \phi_1J_1(\lambda_1e A_0)
\sin k_+\right.\nonumber\\
&&\left.-\sin \phi_2J_1(\lambda_2e A_0)\sin k_-\right].\nonumber 
\end{eqnarray} 
Here, we simplify the notation by introducing the following shorthand:  $k_{\pm}=k_{x}\pm k_{y}$, 
$\lambda_{1}=\sqrt{2-\sin^{2}\theta(\sin2\phi+1)}$, $\lambda_{2}=\sqrt{2+\sin^{2}\theta(\sin2\phi+1)}$, $\lambda_{3}=\sin\theta$, 
$\sin\phi_{1}=(\sin\phi-\cos\phi)/\lambda_{1}$, $\sin\phi_{2}=(\sin\phi+\cos\phi)/\lambda_{2}$, and 
$J_{n}(x)$ denotes Bessel functions
of the first kind [derivation details are provided in Sec. I of the Supplemental Material (SM)~\cite{supplemental}]. 

The positions of FWPs are simply determined by the zeros of the coefficients 
in front of the Pauli matrices. Without loss of generality, we focus on a specific case for a detailed analysis. 
Specifically, we consider light propagating along the direction 
$(\theta,\phi)=(\frac{\pi}{3},\frac{\pi}{3})$. The resulting 
positions and chiralities of FWPs are illustrated in Fig. \ref{fig2}(b). 
As expected, we observe four pairs of FWPs in the $k_{z}=0$ plane 
and another four pairs in the $k_{z}=\pi$ plane. Notably, 
the distributions of positions and chiralities in the two
$k_{z}$ planes are identical. 
Consequently, when the bulk Weyl points are projected 
along the $z$ direction, those with the same topological charge overlap, yielding a net topological charge of
$+2$ or $-2$ per Weyl-point projection. Each Weyl-point projection is thus accompanied 
by two Fermi arcs, a property we will demonstrate below.

\begin{figure}[t]
\centering
\includegraphics[width=0.45\textwidth]{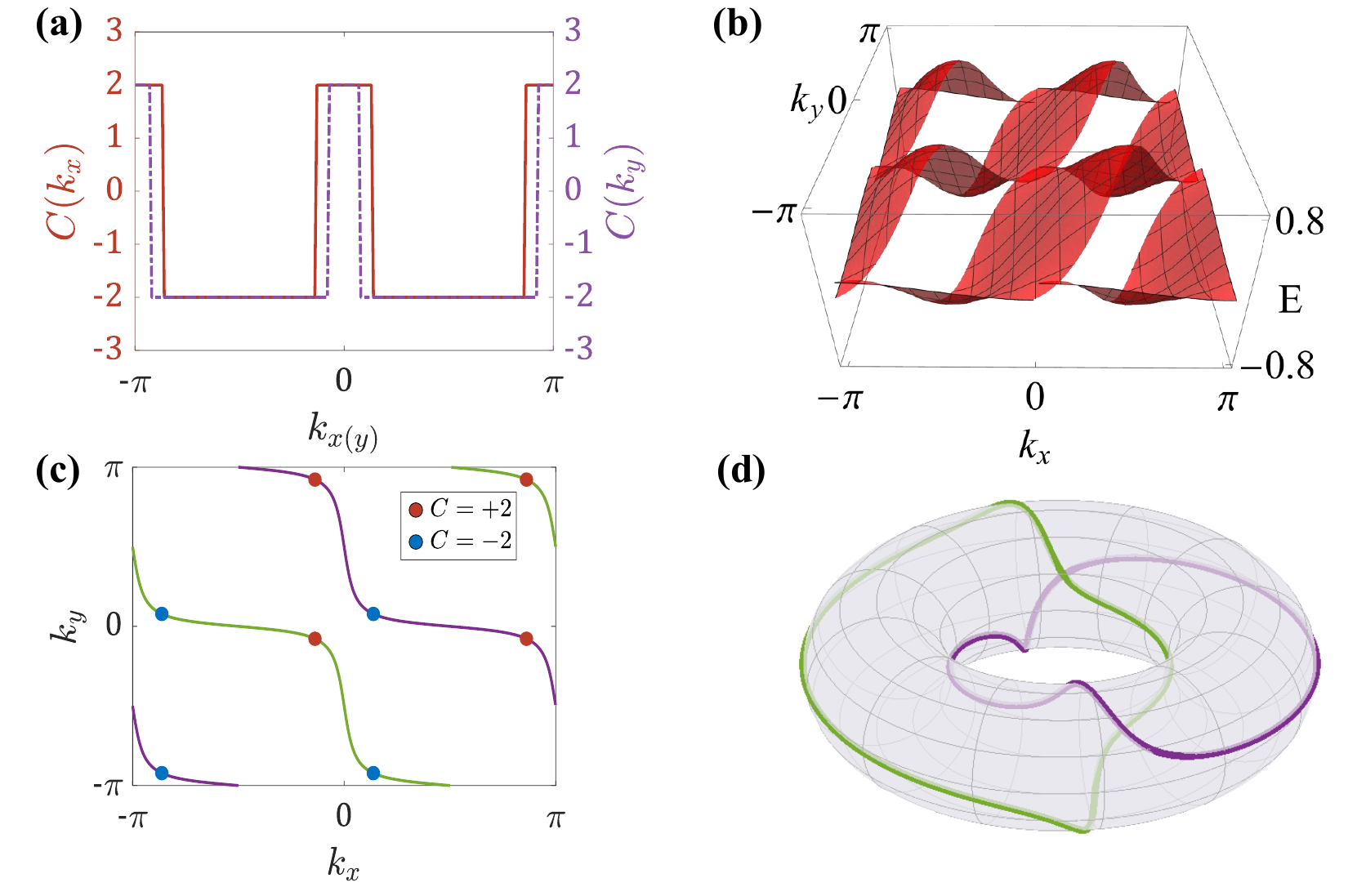}
\caption{ (a) Momentum-dependence of layer Chern numbers. (b) Surface-state 
spectrum on the top $z$-normal surface. (c) Fermi arcs at zero energy (green and purple curves), 
with red and blue markers indicating Weyl-point projections. 
(d) Linked topology of Fermi arcs visualized on the surface BZ torus. 
Parameters are $eA_{0}=0.8$, $\omega=5$, $\eta=1$, $\theta=\pi/3$, and $\phi=\pi/3$. 
}\label{fig3}
\end{figure}

Before identifying the Fermi arcs, we first calculate the momentum-resolved layer Chern number
$C(k_{i})$ for $i\in\{x,y,z\}$. From the Weyl point distribution in Fig.~\ref{fig2}(b), we observe that
$C(k_{z})$ vanishes identically, whereas $C(k_{x})$ and $C(k_{y})$ remain nonzero across their 
respective gapped planes. The explicit momentum dependence of $C(k_{x})$ and $C(k_{y})$ 
is plotted in Fig.~\ref{fig3}(a). Notably, both $|C(k_{x})|$ and $|C(k_{y})|$ take the value $2$,
signifying that under open boundary conditions along the $z$ direction, two chiral Fermi arcs 
emerge for any fixed $k_{x}$ or $k_{y}$. Consequently, the Fermi arcs span the entire surface 
BZ along both $k_{x}$ and $k_{y}$. Intriguingly, $C(k_{x})$ and $C(k_{y})$ 
undergo sign reversal across planes containing Weyl points of identical chirality. 
This behavior demonstrates that the chirality of the Fermi arcs is momentum-dependent, 
flipping sign at Weyl-point projections---a sharp contrast to 3D Chern insulators~\cite{Liu2022Chern,Devescovi2021,Yang2025Chern}, 
where the layer Chern numbers remain constant along a given direction, 
resulting in a surface-state Fermi loop of fixed chirality.
In the present system, we find that a 3D Chern insulator phase can also 
be realized by further increasing the light amplitude to annihilate all FWPs 
(see Sec. III of the SM~\cite{supplemental}).

The unique momentum dependence of $C(k_{x})$ and $C(k_{y})$ indicates that 
the Fermi arcs on the $z$-normal surfaces will exhibit a linked configuration. 
To demonstrate this, we need to determine the spectrum of the surface states.  
This can be done by numerically 
calculating the energy spectrum for a sample with periodic boundary conditions in the $x$ and $y$ directions 
and open boundary conditions in the $z$ direction (see Sec. II of the SM~\cite{supplemental}). Notably, for the current 
Hamiltonian, the surface-state spectra can be analytically obtained as the Floquet Hamiltonian in Eq.~(\ref{latticeH})
takes a form similar to the Su-Schrieffer-Heeger model in the $z$ direction~\cite{Su1979}. Specially, 
the surface-state spectra for the top $(+)$ and bottom $(-)$ $z$-normal surfaces are given by 
\begin{eqnarray}
E_{\pm}(\bk_{s})=\mp \text{sgn}(CD(\bk_{s}))F(\bk_{s}),
\end{eqnarray}
where the function $\text{sgn}(x)$ equals $1$ $(-1)$ for $x>0$ ($x<0$). 
Fig.~\ref{fig3}(b) displays the spectrum for the top surface. By tracing the zero-energy contours 
and Weyl-point projections, we obtain the connectivity of the Fermi arcs, as depicted in Fig.~\ref{fig3}(c).
Indeed, each Weyl-point projection connects to  two Fermi arcs, one from each side.  Interestingly these Fermi arcs 
collectively span the entire surface BZ. 
When the 2D surface BZ is compactified into a torus by identifying opposite boundaries, 
the Fermi arcs clearly organize into two mutually linked loops, as illustrated in Fig.~\ref{fig3}(d).
While Fermi arcs spanning the entire surface BZ have been experimentally observed 
in chiral topological semimetals~\cite{Chang2017chiral,Rao2019,Schroter2019,Takane2019}, the linked structure 
we find here represents a novel feature that has not been previously 
identified in either static or Floquet Weyl semimetals.

{\it Light-induced anomalous Hall effect.---}The emergence of FWPs also leads to the anomalous 
Hall effect. This light-induced anomalous Hall effect 
has been experimentally observed in various Dirac semimetals under CPL irradiation, 
including 2D graphene~\cite{McIver2020} and 3D Co$_{3}$Sn$_{2}$S$_{2}$~\cite{Naotaka2025AHE}.  
Since the system is driven out of equilibrium, we employ the non-equilibrium 
Floquet Green's function method to calculate the Hall current and 
conductivity~\cite{Oka2009,Kitagawa2011Floquet,Mosallanejad2024}.  Specially, we consider a system of size $N=N_{x}N_{y}N_{z}$, 
where $N_{i}$ denotes the number of lattice sites along the $i$-th direction.  
A bias voltage $V$ is applied along the $z$ direction, requiring open boundary conditions in $z$, 
while periodic boundary conditions are imposed along the $x$ and $y$ directions 
for computation simplicity. The coupling between the sample and electrodes is set as constant
$\Gamma$. Here, we focus on the light-induced Hall current 
along the $y$ direction, as the original 
static Hamiltonian possesses $C_{4z}$ rotation symmetry.
The time-average dc Hall current over one period is given by (with $e=\hbar=1$)
\begin{eqnarray}
	J^y=-\frac{i}{N}\sum_{k_x,k_y}\int_0^\omega\frac{d\varepsilon}{2\pi}
	\text{Tr}[G^<_{nm,ij}(\varepsilon,\mathbf{k}_{s})v^y_{mn,ji}(\mathbf{k}_{s})],
\end{eqnarray}
where  $G^<_{nm,ij}$ is the Floquet lesser Green's function and $v^y_{mn,ji}(\mathbf{k}_{s})$ is 
the Floquet velocity matrix elements (see Sec. IV of the SM~\cite{supplemental} for details). The indices $\{n,m\}$ label 
the sector in the frequency space, while $\{i,j\}$ label lattice sites along the $z$ direction.   
Repeated indices are implicitly summed over. Once the current is determined, the Hall conductivity 
$\sigma_{yz}$ is calculated as $\sigma_{yz}=J_{y}/E_{z}=J_{y}N_{z}/V$. 

\begin{figure}[t]
\centering
\includegraphics[width=0.45\textwidth]{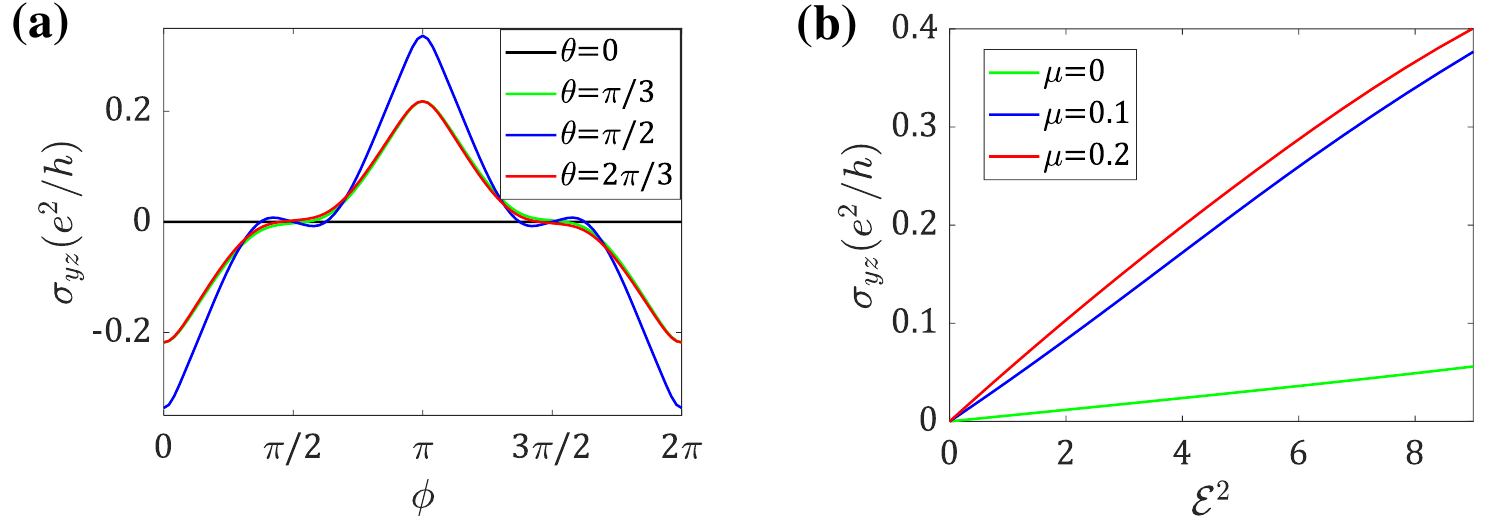}
\caption{ Zero-temperature Hall conductivity. 
(a) Hall conductivity $\sigma_{yz}$ as a function of the light's propagation direction, calculated for $\mu=0$ 
with parameter $eA_{0}=0.8$. 
(b) $\sigma_{yz}$ as a function of the light's intensity, with  
$\theta=\pi/3$, and $\phi=5\pi/6$. Common parameters are $\omega=5$, $\eta=1$, 
$V=0.1$, $\Gamma=0.2$, and $N_{x}\times N_{y}\times N_{z}=40\times40\times10$.
}\label{fig4}
\end{figure}

Figure \ref{fig4}(a) shows the evolution of the Hall conductivity $\sigma_{yz}$
as a function of the light's propagation direction. Notably, $\sigma_{yz}$ vanishes 
identically when $\theta=0$ (and equivalently for $\theta=\pi$), since 
 CPL perpendicular to SNLs cannot induce FWPs and  preserves $\mathcal{PT}$ symmetry in this configuration. 
 Interestingly, $\sigma_{yz}$ also remains  zero at $\phi=\pi/2$ and $\phi=3\pi/2$, 
corresponding to light incident along the angular bisector of the perpendicularly crossed SNLs. 
In these cases, the FWPs are symmetrically distributed, resulting in a complete cancellation of 
their contributions to the Hall conductivity. These findings demonstrate a strong dependence of
$\sigma_{yz}$ on the light's orientation. 
In Fig.~\ref{fig4}(b), we further analyze the evolution of 
$\sigma_{yz}$ with respect to the light intensity, expressed as 
$\mathcal{E}^{2}$, where $\mathcal{E}=\omega A_{0}$ is the electric field amplitude). 
Within the weak intensity regime, the Hall conductivity exhibits a linear dependence 
on the light intensity. This linear behavior is universal regardless of the chemical potential $\mu$. 
However, tuning $\mu$ can strongly affects the slope of the linear dependence, 
as shown in Fig.~\ref{fig4}(b).

{\it Discussions and conclusions.---}We have demonstrated that CPL can 
controllably generate FWPs from intersected SNLs. This discovery 
advances our fundamental understanding of light-driven topological transitions 
between nodal-line semimetals and Weyl semimetals, while revealing a novel 
Weyl semimetal phase characterized by linked Fermi arcs.
Beyond enabling momentum-dependent chiral transport, these linked Fermi arcs are 
expected to host exotic transport phenomena under magnetic fields. This arises because 
the linking structure produces unique Weyl orbitals~\cite{Zhang2021}---the electron’s isoenergy trajectories 
in a magnetic field---as multiple closed paths become available. Consequently, 
we anticipate the emergence of exotic signatures in quantum oscillations~\cite{Potter2014,Moll2016} and 
3D quantum Hall effects~\cite{Wang2017QHE,Zhang2017QHE,zhang2019quantum} tied to these Weyl orbitals.

{\zb Although Weyl points are robust against perturbations and long Fermi arcs can appear 
on general surfaces, observing the arcs' linking structure requires 
careful sample surface preparation. In our model, the linking structure 
observed on $z$-normal surfaces arises from two closely-related properties of the Hamiltonian: the overlap of  
projections of bulk Weyl points with identical 
topological charges, and layer Chern numbers $|C(k_{x})|$ and $|C(k_{y})|$
 that are identically 2 across all gapped planes. However, this linking structure 
 is not robust against the variation of surface orientation. For a general surface orientation, 
 the Weyl-point projections no longer overlap, and the layer Chern numbers vary across the Brillouin zone, causing the linking to disappear. 
 If the cleavage surface only slightly deviates from the $z$-normal direction, 
 the Fermi arcs remain proximate but unlinked. Under such conditions, the essential physics of 
 multiple Weyl orbitals persists. In particular, the effect known as magnetic breakdown can still enable 
 electrons to tunnel between adjacent Weyl-orbital trajectories~\cite{Alexandradinata2018}.}

Intersected SNLs can appear in both nonmagnetic and magnetic materials. Here we 
highlight the relevance of our theoretical predictions to spin-split antiferromagnets---a class of 
materials exhibiting momentum-dependent spin splitting and symmetry-enforced 
zero-net magnetization~~\cite{Hayami2019AM,Hayami2020AM,Yuan2020AM,Yuan2021AM,
Liu2022AMPRX,Ma2021AM,Libor2022AMa,Libor2022AMb,Libor2022AMc}. The underlying symmetries responsible for these characteristics 
naturally  stabilize  SNLs along high symmetry lines in the BZ~\cite{Fernandes2024,Zhuang2025SNL}. Notably, 
recent theoretical work has identified several existing compounds---including 
$\beta$-Fe$_{2}$(PO$_{4}$)O, Co$_{2}$(PO$_{4}$)O 
and LiTi$_{2}$O$_{4}$---as altermagnets 
hosting such intersected SNLs~\cite{He2024SNL}. Based on our study, a light-induced topological transition from 
SNLs to Weyl points can be probed by TrARPES in these materials. {\zb We now provide an estimation to
the distance of two neighboring FWPs. Specifically, we 
consider a mid-infrared laser with frequency $\hbar\omega=280$ meV (here the reduced Planck constant is restored) and 
amplitude  $\mathcal{E}=\omega A_{0}=2.1\times 10^{8}$ V/m~\cite{Aeschlimann2021}. 
Assuming that the laser is incident along the direction $(\theta,\phi)=(\pi/2,\pi/4)$, 
we obtain $d_{\omega}=\sqrt{2}eA_{0}/\hbar=\sqrt{2}e\mathcal{E}/\hbar\omega\simeq 1\times 10^{9}$ m$^{-1}$.
For the material candidate $\beta$-Fe$_{2}$(PO$_{4}$)O (in-plane lattice constant 
 $a=5.419\times10^{-10}$ m)~\cite{He2024SNL}, this distance corresponds to roughly a tenth of the Brillouin-zone size
($G=2\pi/a=1.2\times 10^{10}$ m$^{-1}$). 
This light-induced separation is several times larger than that observed in TaAs~\cite{Lv2015weyl,Xu2015science}, 
confirming its experimental viability.}

Beyond quantum materials, the predicted Weyl phase characterized by linked Fermi arcs 
also offers exciting opportunities for investigating unconventional 
bulk-boundary physics  in classical systems~\cite{rechtsman2013photonic,Fleury2016,Ozawa2019review,Ma2019review,Zhu2023review,Wang2023hopf}.
The minimal two-band Hamiltonian in Eq.~(\ref{latticeH}) also serves as 
an ideal model for experimental exploration due to its simplicity, 
enabling direct implementation in classical systems.

{\it Acknowledgements.---}This work is supported by the National Natural Science Foundation of China (Grant No. 12174455), 
and Guangdong Basic and Applied Basic Research Foundation (Grant No. 2023B1515040023).

\bibliography{floquet.bib}

\end{document}